# Multi-observable Uncertainty Equality based on the sum of standard deviations in the qubit system

Xiao Zheng(郑晓)[1], Shaoqiang Ma(马少强)[1], Guofeng Zhang(张国锋)[1] *

[1]*Key Laboratory of Micro-Nano Measurement-Manipulation and Physics (Ministry of Education), School of Physics and Nuclear Energy Engineering; State Key Laboratory of Software Development Environment, Beihang University, Xueyuan Road No. 37, Beijing 100191, China*

**Abstract:** We construct a multi-observable uncertainty *equality* as well as an inequality based on the sum of standard deviations in the qubit system. The obtained equality indicates that the uncertainty relation can be expressed more accurately, and also can be used to detect the mixedness of the system. Meanwhile, the new uncertainty inequality can provide a tighter lower bound, and the tightness can be maintained at a high level even in an open system. Furthermore, the deficiency of the uncertainty relation, that the lower bound of the product form uncertainty relations can be null even for two incompatible observables, can be completely fixed by the new uncertainty relation.

## I. Introduction

Quantum uncertainty relations are one of the most fundamental differences between quantum and classical mechanics [1-3]. Any pair of incompatible observables admits a certain form of uncertainty relation [4-8], which expresses the impossibility of the jointly sharp preparation of these incompatible observables [9-12]. The uncertainty relation has been widely used in the quantum information science, such as entanglement detection [1,13-15], quantum squeezing [16-19], and quantum metrology [20-22]. The initial investigation of the uncertainty relation was mainly focused on the product form, such as the Schrödinger uncertainty relation (SUR) [3]:

$$\Delta A^2 \Delta B^2 \geq \left|\frac{1}{2i}\langle [A,B]\rangle\right|^2 + \left|\frac{1}{2}\langle \{\breve{A},\breve{B}\}\rangle\right|^2 \qquad (1)$$

where the variance $\Delta O^2$ and expected value $\langle O \rangle$ are calculated on the state $\rho$, and $\breve{O} = O - \langle O \rangle$. However, the product form uncertainty relations cannot fully capture the concept of incompatible observables because the lower bounds of them may be null even for two incompatible observables. This deficiency is referred to as the triviality problem of the uncertainty relation. In order to fix the

---

* Correspondence and requests for materials should be addressed to G.Z.(gf1978zhang@buaa.edu.cn)

triviality problem, the uncertainty relation based on the sum of variances have been investigated [23]:

$$\Delta A^2 + \Delta B^2 \geq |\langle\psi|A \pm iB|\psi^\perp\rangle|^2 \pm i\langle[A,B]\rangle, \qquad (2)$$

where $|\psi^\perp\rangle$ is the state orthogonal to the state of the system $|\psi\rangle$. The lower bound (2) can be guaranteed to be non-zero for incompatible observables. Thus the triviality problem can be fixed by the sum uncertainty relation (2). However, due to the existence of $|\psi^\perp\rangle$, the lower bound (2) is difficult to be extended to the mixed state [23].

In addition to the uncertainty relations for two incompatible observables, there exist the ones for three or more incompatible observables. The multi-observable uncertainty relations have a wider application than the one for two incompatible observables. The recently derived uncertainty relations for three or N observables are [24, 25]:

$$\Delta A_1^2 + \Delta A_2^2 + \Delta A_3^2 \geq \frac{1}{3}\Delta(A_1+A_2+A_3)^2 + \frac{\sqrt{3}}{3}|\langle[A_1,A_2,A_3]\rangle| \qquad (3)$$

$$\Delta A_1^2 + \Delta A_2^2 + \Delta A_3^2 \geq \frac{\sqrt{3}}{3}\{|\langle[A_1,A_2]\rangle| + |\langle[A_2,A_3]\rangle| + |\langle[A_3,A_1]\rangle|\} \qquad (4)$$

$$\sum_{m=1}^{N}\Delta A_m^2 \geq \frac{1}{N-2}\sum_{1\leq m<n\leq N}\Delta(A_m+A_n)^2 - \frac{1}{(N-1)^2(N-2)}[\sum_{1\leq m<n\leq N}\Delta(A_m+A_n)]^2 \qquad (5)$$

where $[A_1,A_2,A_3] = [A_1,A_2] + [A_2,A_3] + [A_3,A_1]$.

In this paper, we construct a multi-observable uncertainty *equality* as well as an inequality based on the sum of standard deviations in the qubit system. The outline of the paper is as follows. In section II, the uncertainty equality as well as an inequality based on the sum of standard deviations is formulated in the qubit system. We demonstrate that the new uncertainty inequality is tighter than other recent uncertainty relations, and the tightness can be maintained at a high level even in the open system. Then, a proof, that the triviality problem of the product form uncertainty relation can be completely fixed by the obtained uncertainty relation, is presented in section III. In section IV, we show that the uncertainty equality can be used as a measure of the mixedness of the system which usually is expensive in terms of resources involved. Finally, the last section is devoted to the discussion and conclusion.

## II. Uncertainty Equality Based on the Sum of Standard Deviations

Consider N arbitrary Hermitian operators $A_1, A_2, \ldots, A_N$ in the qubit system, and the uncertainties of the corresponding outcomes when they are measured admit the following uncertainty equality:

$$\sum_{m=1}^{N}\Delta A_m = \frac{1}{2(N-1)}\sum_{i,j=1,i\neq j}^{N}[2\sqrt{M\mathcal{F}([A_i,A_j]) + |\mathcal{G}(A_i,A_j)|^2} + \Delta_{cms}^2 A_i + \Delta_{cms}^2 A_j -$$

$$\zeta(A_i, A_j)]^{1/2} \tag{6}$$

where $\mathcal{F}([A_i, A_j]) = \text{tr}\left([A_i, A_j][A_i, A_j]^\dagger\right)/4$, $\mathcal{G}(A_i, A_j) = \langle A_i \cdot A_j \rangle - \langle A_i \rangle \langle A_j \rangle$, and $\zeta(A_i, A_j) = (\langle A_i \rangle - \langle A_i \rangle_{cms})^2 + (\langle A_j \rangle - \langle A_j \rangle_{cms})^2$. $M = 1 - \text{tr}(\rho^2)$ is the mixedness of the state $\rho$ [26], and the mixedness is equal to zero for pure state and greater than zero for mixed state. The variance $\Delta^2_{cms}A_i$ and the expected value $\langle A_i \rangle_{cms}$ are calculated on the completely mixed state $\rho_{cms} = I/2$, namely the state which possesses the maximum mixedness [26], with $I$ being the identity matrix. Based on the definition of $\mathcal{F}([A_i, A_j])$, $\mathcal{F}([A_i, A_j])$ can be rewritten as $\|[A_i, A_j]\|/4$, namely the modulus of commutator $[A_i, A_j]$, with $\|O\|$ being defined as $\text{tr}(OO^\dagger)$. Thus, $\mathcal{F}([A_i, A_j])$ is used to quantify the noncommutativity between $A_i$ and $A_j$. $\mathcal{G}(A_i, A_j) = \langle A_i \cdot A_j \rangle - \langle A_i \rangle \langle A_j \rangle$ is the covariance between $A_i$ and $A_j$, and can thus be used to measure the correlation between $A_i$ and $A_j$. $(\langle A_i \rangle - \langle A_i \rangle_{cms})^2$ is the difference between the expected value of $A_i$ on the state of the system and the expected value of $A_i$ on the completely mixed state, and $\zeta(A_i, A_j)$ thus represents the sum of the differences related to $A_i$ and $A_j$.

*Proof*: In Bloch sphere representation, the density matrix of the qubit system can be expressed as [27, 28]:

$$\rho = \frac{1}{2}(I + p_1 \sigma_x + p_2 \sigma_y + p_3 \sigma_z), \tag{7}$$

where $\sigma_x, \sigma_y, \sigma_z$ are standard Pauli matrices and $p_1, p_2, p_3$ denote real parameters with $p_1^2 + p_2^2 + p_3^2 \leq 1$. Additionally, an arbitrary two-dimension Hermitian operator can be written as a linear combination of $\{\sigma_x, \sigma_y, \sigma_z, I\}$:

$$A_m = a_{m1}\sigma_x + a_{m2}\sigma_y + a_{m3}\sigma_z + a_{m4}I. \tag{8}$$

where $a_{mi}$ is real parameters ($i = 1,2,3,4$). Based on the assumption above, we have:

$$\Delta A_m = \{\sum_{i=1}^{3}(1 - p_i^2)a_{mi}^2 - 2[p_2 p_3 a_{m2} a_{m3} + p_1 a_{m1}(p_2 a_{m2} + p_3 a_{m3})]\}^{1/2} \tag{9}$$

$$\mathcal{F}([A_m, A_n]) = 2[(a_{m3}a_{n2} - a_{m2}a_{n3})^2 + (a_{m2}a_{n1} - a_{m1}a_{n2})^2 + (a_{m3}a_{n1} - a_{m1}a_{n3})^2] \tag{10}$$

$$M = \frac{1}{2}(1 - p_1^2 - p_2^2 - p_3^2) \tag{11}$$

$$\Delta^2_{cms}A_m = a_{m1}^2 + a_{m2}^2 + a_{m3}^2 \tag{12}$$

$$\zeta(A_m, A_n) = (\sum_{i=1}^{3} p_i a_{mi})^2 + (\sum_{i=1}^{3} p_i a_{ni})^2 \tag{13}$$

Here, the analytic expression of $\mathcal{G}(A_m, A_n)$, due to its complex form, will not be presented. Combining above formulas, one can obtain uncertainty equality (6), and therefore the proof is completed. The uncertainty equality indicates that the uncertainty relation can be expressed exactly in the qubit system.

Based on the definition of $\mathcal{F}([A_i, A_j]) = \|[A_i, A_j]\|/4$, we can see that $\mathcal{F}([A_i, A_j])$ is in fact a state-independent measure to quantify the noncommutativity between $A_i$ and $A_j$. To investigate the effect of such a state-independent noncommutativity measure on the uncertainty relation, we eliminate the non-negative and state-independent term $|\mathcal{G}(A_m, A_n)|^2$ in the uncertainty equality (6). Based on the uncertainty equality (6) and using $|\mathcal{G}(A_m, A_n)|^2 \geq 0$, one can obtain a multi-observable uncertainty inequality based on the sum form of standard deviations:

$$\sum_{m=1}^{N} \Delta A_m \geq \frac{1}{2(N-1)} \sum_{i,j=1, i \neq j}^{N} [2\sqrt{M\mathcal{F}([A_i, A_j])} + \Delta_{cms}^2 A_i + \Delta_{cms}^2 A_j - \zeta(A_i, A_j)]^{1/2} \quad (14)$$

The uncertainty equality (6) is saturated for any qubit state, and thus we mainly focus on the performance of the new uncertainty inequality (14).

The comparison between the new uncertainty relation (14) and other recent ones will be presented in the following. It turns out to be a relatively reasonable way to compare different types of uncertainty relations by their tightness, which is defined as dividing both sides of the inequalities by their own lower bound [28]. Thus the tightness of (14) (3) (4) and (5) are obtained as:

$$T1 = \frac{\sum_{m=1}^{N} \Delta A_m}{\frac{1}{2(N-1)} \sum_{i,j=1, i \neq j}^{N} [2\sqrt{M\mathcal{F}([A_i, A_j])} + \Delta_{cms}^2 A_i + \Delta_{cms}^2 A_j - \zeta(A_i, A_j)]^{1/2}} \quad (15)$$

$$T2 = \frac{\Delta A_1^2 + \Delta A_2^2 + \Delta A_3^2}{\frac{1}{3}\Delta(A_1 + A_2 + A_3)^2 + \frac{\sqrt{3}}{3}|\langle[A_1, A_2, A_3]\rangle|} \quad (16)$$

$$T3 = \frac{\Delta A_1^2 + \Delta A_2^2 + \Delta A_3^2}{\frac{\sqrt{3}}{3}\{|\langle[A_1, A_2]\rangle| + |\langle[A_2, A_3]\rangle| + |\langle[A_3, A_1]\rangle|\}} \quad (17)$$

$$T4 = \frac{\sum_{m=1}^{N} \Delta A_m^2}{\frac{1}{N-2} \sum_{1 \leq m < n \leq N} \Delta(A_m + A_n)^2 - \frac{1}{(N-1)^2(N-2)} [\sum_{1 \leq m < n \leq N} \Delta(A_m + A_n)]^2} \quad (18)$$

As shown in Fig.1, we can see that uncertainty relation (14) is tighter than the other three ones.

In the open system, the ubiquitous interaction with the environment inevitably deduces the increasing of the mixedness of the system [29], and thus it becomes important to investigate the performance of the uncertainty relations in the mixed states. However, as shown in Fig.2, the tightness of the uncertainty relations (3), (4), and (5) becomes worse and worse with the mixedness increasing. In other words, the tightness of traditional uncertainty relations performed poorly in the open system. As shown in Fig.2, as the mixedness increases, the tightness of uncertainty relation (14) becomes better and better with. That is to say, the tightness of the new uncertainty inequality can be maintained at a high level even in an open system.

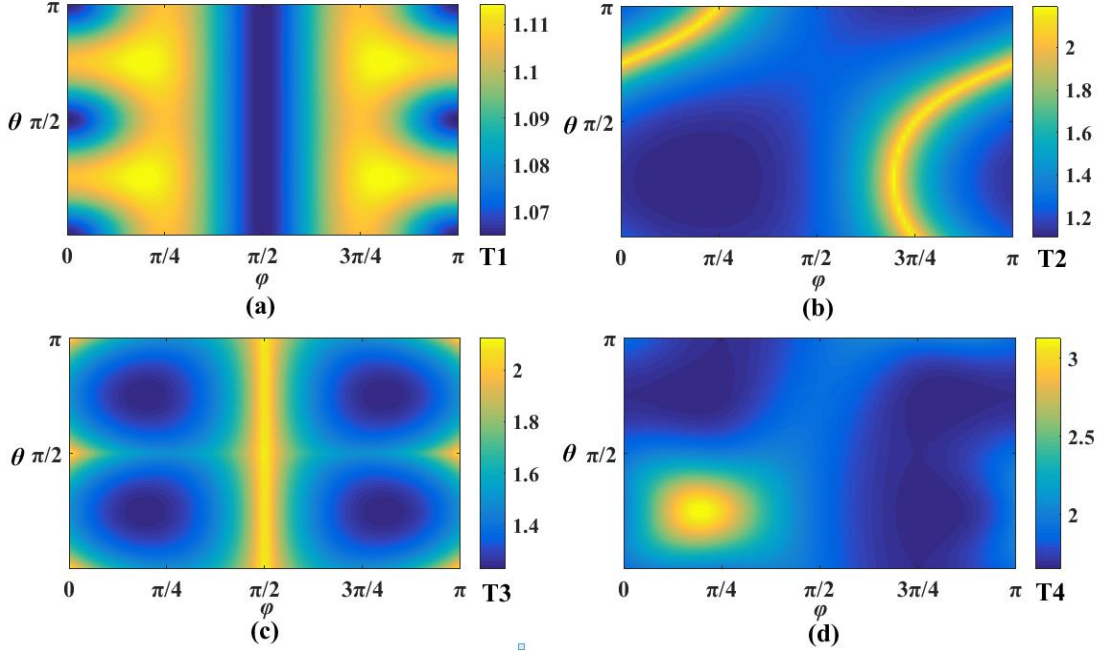

Fig.1: Evolution of the tightness T1, T2, T3, and T4 with respect to the state are shown in the (a), (b), (c) and (d), respectively. We take $A_1 = \sigma_x, A_2 = \sigma_y, A_3 = \sigma_z$, and the state of the system is parameterized by $(\gamma, \theta, \varphi)$ as $\rho = 1/2[\gamma \cos\varphi \sin\theta\, \sigma_x + \gamma\varphi \cos(\theta)\, \sigma_y + \gamma \sin(\varphi)\, \sigma_z + I]$, where $\gamma \in [0,1]$, and $\varphi, \theta \in [0,\pi]$. Here $\gamma = \sqrt{0.8}$, and then $M = (1-\gamma^2)/2 = 0.1$.

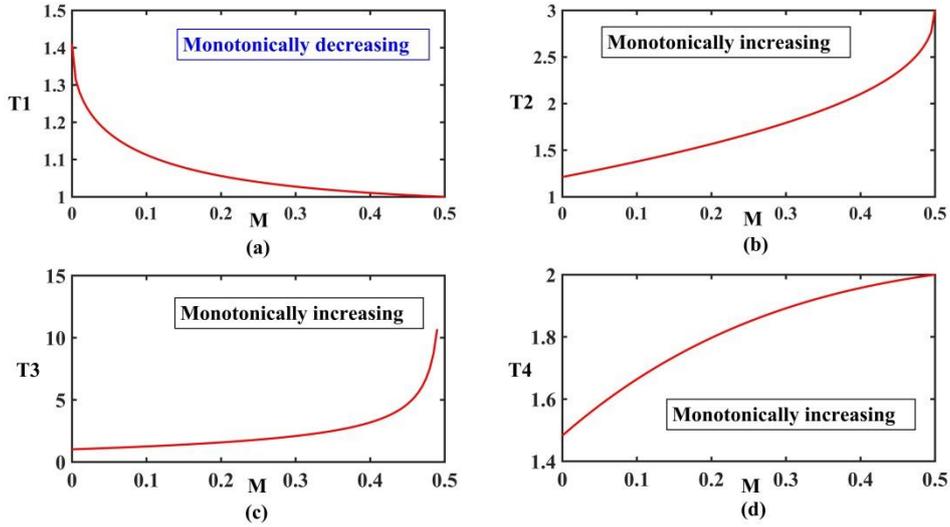

Fig.2: Evolution of the tightness T1, T2, T3, and T4 with respect to the mixedness M of the system are shown in the (a), (b), (c) and (d), respectively. We take $A_1 = \sigma_x, A_2 = \sigma_y, A_3 = \sigma_z$, and the state of the system is parameterized by $(\gamma, \theta, \varphi)$ as $\rho = 1/2[\gamma \cos\varphi \sin\theta\, \sigma_x + \gamma\varphi \cos(\theta)\, \sigma_y + \gamma \sin(\varphi)\, \sigma_z + I]$, where $\gamma \in [0,1]$, $\varphi, \theta \in [0,\pi]$. The mixedness of the system can be obtained as $M = (1-\gamma^2)/2$, and we take $\theta = 3\pi/4$ and $\varphi = \pi/4$.

### III. Fixing the triviality problem

The product form uncertainty relation cannot fully capture the concept of the incompatible observables, because its lower bound can be null even for incompatible observables [23]. Then we will demonstrate that this triviality problem can be completely fixed by the new uncertainty relation. Take N=2, and the new uncertainty relation (14) then turns into:

$$\Delta A + \Delta B \geq \left[2\sqrt{M\mathcal{F}([A,B])} + \Delta_{cms}^2 A + \Delta_{cms}^2 B - \zeta(A,B)\right]^{\frac{1}{2}} \quad (19)$$

where $A = a_1\sigma_x + a_2\sigma_y + a_3\sigma_z + a_4 I$ and $B = b_1\sigma_x + b_2\sigma_y + b_3\sigma_z + b_4 I$ stand two arbitrary observables with $a_i, b_i \in R$. Then, we can deduce:

$$\Delta_{cms}^2 A + \Delta_{cms}^2 B - \zeta(A,B)$$
$$= \sum_{i=1}^{3}(a_i^2 + b_i^2) - (p_x a_1 + p_y a_2 + p_z a_3)^2 - (p_x b_1 + p_y b_2 + p_z b_3)^2$$
$$\geq \sum_i(a_i^2 + b_i^2) - (p_x^2 + p_y^2 + p_z^2)[\sum_i(a_i^2 + b_i^2)]$$
$$= [1 - (p_x^2 + p_y^2 + p_z^2)][\sum_i(a_i^2 + b_i^2)] \geq 0 \quad (20)$$

where $p_x^2 + p_y^2 + p_z^2 \leq 1$ is used. It worth mentioning that the necessary condition for $\Delta_{cms}^2 A + \Delta_{cms}^2 B - \zeta(A,B) = \sum_i(a_i^2 + b_i^2) - (p_x^2 + p_y^2 + p_z^2)[\sum_i(a_i^2 + b_i^2)]$ is $a_1/b_1 = a_2/b_2 = a_3/b_3$, which indicates $[A,B] = 0$. Thus, we have $\Delta_{cms}^2 A + \Delta_{cms}^2 B - \zeta(A,B) > 0$ when $[A,B] \neq 0$. That is to say, the triviality problem can be completely fixed by the new uncertainty relation (14). For instance, we take $A = \sigma_x$, $B = \sigma_y$ and a set of states parameterized by θ as $|\psi\rangle = \cos(\theta)|1\rangle + \sin(\theta)|0\rangle$ with $|1\rangle$ and $|0\rangle$ being the eigenstates of $\sigma_z$. Obviously, $[A,B] \neq 0$ and there exists no common eigenstate between A and B. Then, one can obtain that:

$$L_{new} = \sqrt{1 + \cos(2\theta)^2} \quad (21)$$
$$L_{SUR} = \cos(2\theta)^2 \quad (22)$$

where $L_{new} = [2\sqrt{M\mathcal{F}([A,B])} + \Delta_{cms}^2 A + \Delta_{cms}^2 B - \zeta(A,B)]^{1/2}$ is the lower bound of the new uncertainty inequality (19), and $L_{SUR} = |\langle[A,B]\rangle|^2/4 + |\langle\{\check{A},\check{B}\}\rangle|^2/4$ is the lower bound of the SUR. The evolutions of the two lower bounds are shown in Fig.3. It can be seen that SUR will be trivial for $\theta = \pi/4$ and $3\pi/4$, and the triviality problem can be easily fixed by the new uncertainty inequality.

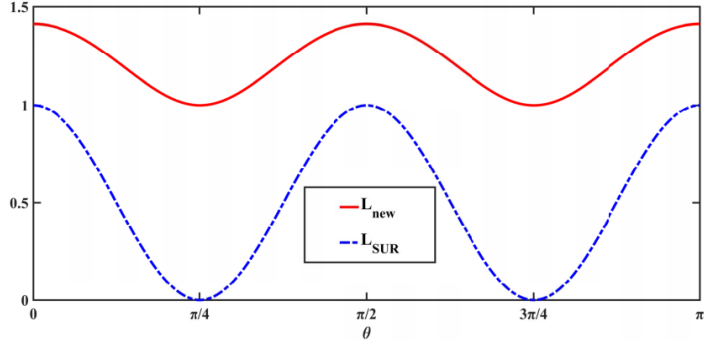

Fig.3: Evolutions of $L_{new}$ and $L_{SUR}$ with respect to θ.

**IV. Detection of the mixedness**

As mentioned above, the ubiquitous interaction with the environment inevitably affects the purity of a quantum system at the practical level, and it becomes important for an experimenter to test the mixedness of the system so as to use it effectively as a resource for the quantum information processing [29]. But in the practical applications, the mixedness of the system usually is impossible to be obtained by direct calculations and is expensive in term of the physical resources and measurements involved [27, 29]. Therefore, it is become meaningful to find an effective method to detect the mixedness of the system [29]. The obtained uncertainty equality can be used to measure the mixedness of the system. Take N=2, and then the expression of the mixedness can be obtained by a deformation of Eq. (6):

$$M = \frac{\left((\Delta A+\Delta B)^2 - \Delta^2_{cms}A - \Delta^2_{cms}B + \zeta(A,B)\right)^2 - 4|\mathcal{G}(A,B)|^2}{4\mathcal{F}([A,B])} \quad (23)$$

The terms in the right hand of the Eq. (23) can be divided into four categories: (i) the operator function $\Delta^2_{cms}A$, $\Delta^2_{cms}B$ and $\mathcal{F}([A,B])$; (ii) the function of expectation $\zeta(A,B)$; (iii) the standard deviation $\Delta A$ and $\Delta B$; (iv) the covariance $\mathcal{G}(A,B)$. The value of the operator function, which has no relationship with the system, can be obtained by direct calculations. The expectation, standard deviation and covariance can be obtained by measuring the related observables and observing the expectation, standard deviation as well as covariance of the measurement result, such as $\Delta A$ can be obtained by making measurement $A$ on the system and observing the standard deviation of the measurement result. Thus the mixedness of the system can be easily obtained by detecting the expectation, standard deviation and covariance involved.

**V. Conclusion**

In conclusion, we have constructed a multi-observable uncertainty *equality* as well as an inequality based on the sum of standard deviations in the qubit system. The uncertainty equality indicates that the uncertainty relation can be expressed exactly. Meanwhile, the obtained uncertainty equality can also be used as a measure of the mixedness which usually is expensive in terms of resources involved. As for the uncertainty inequality, we demonstrate that the new uncertainty inequality is tighter than other uncertainty relations, and the tightness can be maintained as a high level in the mixed state, which means the new uncertainty relation has a good performance even in the open system. Furthermore, we prove that the deficiency in the product form uncertainty relation can be completely fixed by the new uncertainty inequality.

**Acknowledgments**

This work is supported by the National Natural Science Foundation of China (Grant No. 11574022 and 11174024), and the Open Project Program of State Key Laboratory of Low-Dimensional Quantum Physics (Tsinghua University) grants Nos. KF201407, also supported by the Open Project Program of State Key Laboratory of Theoretical Physics, Institute of Theoretical Physics, Chinese Academy of Sciences, China (No.Y4KF201CJ1) and Beijing Higher Education (Young Elite Teacher Project) YETP 1141.

**Reference:**

[1] Heisenberg, W. Über den anschaulichen Inhalt derquantentheoretischen Kinematik und Mechanik. *Z. Phys.* **43**, 172 (1927).

[2] Robertson, H. P. The uncertainty principle. *Phys. Rev.* **34**, 163 (1929).

[3] Schrodinger, E. Sitzungsberichte der Preussischen Akademie der Wissenschaften. *Physikalisch-mathematische Klasse* **14**, 296 (1930).

[4] Kraus, K. Complementary observables and uncertainty relations. *Phys. Rev. D* **35**, 3070 (1987).

[5] Zhang, G. F. Sudden death of entanglement of a quantum model. *Chinese Physics* **16**, 1855 (2007).

[6] Berta, M. Christandl, M. Colbeck, R. Renes, J. M. and Renner, R. The uncertainty principle in the presence of quantum memory. *Nat. Phys.* **6**, 659 (2010).

[7] Oppenheim, J. and Wehner, S. The uncertainty principle determines the nonlocality of quantum mechanics. *Science* **330**, 1072 (2010).

[8] Li, C.F. Xu, J. S. Xu, X.Y. Li, K. and Guo, G.C. Experimental investigation of the entanglement-assisted entropic uncertainty principle. *Nat. Phys.* **7**, 752 (2011).

[9] Wehner, S. and Winter, A. Entropic uncertainty relations—A survey. *New J. Phys.* **12**, 025009 (2010).

[10] Tomamichel, M. Renner, R. Uncertainty relation for smooth entropies. *Phys. Rev. Lett.* **106**, 110506 (2011).

[11] Coles, P. J. Colbeck, R. Yu, L. and Zwolak, M. Uncertainty relations from simple entropic properties, *Phys. Rev. Lett.* **108**, 210405 (2012).

[12] Xu, Z. Y. Yang, W. L. and Feng, M. Quantum-memory-assisted entropic uncertainty relation under noise. *Phys. Rev. A* **86**, 012113(2012).

[13] Hofmann, H. F. and Takeuchi, S. Violation of local uncertainty relations as a signature of entanglement. *Phys. Rev. A* **68**, 032103 (2003).

[14] Guhne, O. Characterizing entanglement via uncertainty relations. *Phys. Rev. Lett.* **92**, 117903 (2004).

[15] Schwonnek, R. Dammeier, L. and Werner, R. F, State-Independent Uncertainty Relations and Entanglement Detection in Noisy Systems, *Phys. Rev. Lett.* **119**, 170404 (2017).

[16] Walls, D. F. and Zoller, P. Reduced quantum fluctuations in resonance fluorescence. *Phys. Rev. Lett.* 47,709 (1981).

[17] Wineland, D. J. Bollinger, J. J. Itano, W. M. Moore, F. L. and Heinzen, D. J. Spin squeezing and reduced quantum noise in spectroscopym. *Phys. Rev. A* **46**, R6797 (1992).

[18] Kitagawa, M. and Ueda, M. Squeezed spin states. *Phys. Rev. A* **47**, 5138 (1993).

[19] Ma, J. Wang, X. G. Sun, C. P. and Nori, F. Quantum spin squeezing. *Phys. Rep*. **509**, 89 (2011).

[20] Zhang, G. F. and Li, S. S. Entanglement in a spin-one spin chain. *Solid State Communications* **138**,17(2006).

[21] Giovannetti, V. Lloyd, S. and Maccone, L. Quantum metrology. *Phys. Rev. Lett.* **96**, 010401 (2006).

[22] Giovannetti, V. Lloyd, S. and Maccone, L. Advances in quantum metrology. *Nat. Photonics* **5**, 222 (2011).

[23] Maccone, L. and Pati, A. K. Stronger uncertainty relations for all incompatible observables. *Phys. Rev. Lett.* **113**, 260401 (2014).

[24] Song, Q. C. and Qiao, C. F. Stronger Schrödinger-like uncertainty relations. *Phys. Lett. A* **380**, 2925 (2016).

[25] Chen, B. and Fei, S. M. Sum uncertainty relations for arbitrary N incompatible observables. *Scientific Reports* **5**, 14238 (2015)

[26] Zheng, **X.** and Zhang, G. F. The effects of mixedness and entanglement on the properties of the entropic uncertainty in Heisenberg model with Dzyaloshinski-Moriya interaction. *Quantum Information Processing* **16**, 1(2017).

[27] Zheng, **X.** and Zhang, G. F. Variance-based uncertainty relation for incompatible observers. *Quantum Information Processing* **16**, 167(2017).

[28] Yao, Y. Xiao, X. Wang, X. G. and Sun, C. P. Implications and applications of the variance-based uncertainty equalities. *Phys. Rev. A* **91**, 062113 (2015).

[29] Mal, S. Pramanik, T. Majumdar1, A. S. Detecting mixedness of qutrit systems using the uncertainty relation. *Phys. Rev. A* **87**, 012105 (2013).